\documentclass[12pt]{article}
\usepackage{mathrsfs}
\usepackage{natbib,graphicx,subfigure,setspace,lscape,longtable}
\usepackage{natbib,epsfig,graphicx,bm,bbm}
\usepackage{mathrsfs,amsmath,amsthm,amssymb,color}

\usepackage{indentfirst}

\bibpunct{(}{)}{;}{a}{,}{,}

\newtheorem{theorem}{Theorem}
\newtheorem{lemma}{Lemma}

\newtheorem{proposition}{Proposition}
\newcommand{\csection}[1]
    {\begin{center}
        \stepcounter{section}
        {\bf\large\arabic{section}. #1}
    \end{center}
    \vspace{-0.15 cm}
}

\newcommand{\csubsection}[1]{
\vspace{-0.25 cm}
\begin{center}
\stepcounter{subsection}
{\it\arabic{section}.\arabic{subsection}. #1}
\end{center}
\vspace{-0.25 cm}
}

\def\beq{\begin{equation}}
\def\eeq{\end{equation}}
\def\beqr{\begin{eqnarray}}
\def\eeqr{\end{eqnarray}}
\def\beqrs{\begin{eqnarray*}}
\def\eeqrs{\end{eqnarray*}}
\def\bet{\begin{theorem}}
\def\eet{\end{theorem}}
\def\bel{\begin{lemma}}
\def\eel{\end{lemma}}
\def\bep{\begin{proposition}}
\def\eep{\end{proposition}}
\def\bg{\begin{figure}[tbph]\begin{center}}
\def\eg{\end{center}\end{figure}}

\def\bc{\begin{center}}
\def\ec{\end{center}}

\def\n{\nonumber}

\def\beq{\begin{equation}}
\def\eeq{\end{equation}}
\def\beqr{\begin{eqnarray}}
\def\eeqr{\end{eqnarray}}
\def\beqrs{\begin{eqnarray*}}
\def\eeqrs{\end{eqnarray*}}
\def\bet{\begin{theorem}}
\def\eet{\end{theorem}}
\def\bel{\begin{lemma}}
\def\eel{\end{lemma}}
\def\bep{\begin{proposition}}
\def\eep{\end{proposition}}
\def\bg{\begin{figure}[tbph]\begin{center}}
\def\eg{\end{center}\end{figure}}

\def\bc{\begin{center}}
\def\ec{\end{center}}

\def\n{\nonumber}

\def\wh{\widehat}

\def\mB{\mathcal B}

\def\mR{\mathbb{R}}

\def\mS{\mathcal S}

\def\mR{\mathbb{R}}
\def\mS{\mathcal S}

\def\var{\mbox{var}}

\numberwithin{equation}{section}

\newcommand{\ve}{{\varepsilon}}
\renewcommand{\epsilon}{{\ve}}
\renewcommand{\hat}{\widehat}

\numberwithin{equation}{section}

\begin{document}

\begin{center}
{\bf\Large  Hyperparameter Selection for Subsampling Bootstraps}\\
\bigskip
{Yingying Ma$^1$  and Hansheng Wang$^2$}

{\it$^1$School of Economics and Management, Beihang University\\
$^2$Guanghua School of Management, Peking University}
\end{center}

\begin{abstract}

Massive data analysis becomes  increasingly
prevalent, subsampling methods  like   BLB (Bag
of Little Bootstraps) serves as powerful tools  for  assessing the quality of estimators for massive data.  However, the performance of the subsampling methods are highly influenced by the selection of tuning parameters  ( e.g., the subset size, number of resamples per subset ). In this article we develop  a hyperparameter selection methodology, which can be used  to select tuning parameters  for subsampling methods.
Specifically,  by a careful theoretical analysis, we find an   analytically simple and elegant relationship between  the asymptotic efficiency of  various subsampling estimators  and their hyperparameters.
 This leads to an optimal choice of the hyperparameters.  More specifically, for an arbitrarily specified hyperparameter set, we can improve it to be a new set of hyperparameters with  no extra CPU time cost
 but the resulting estimator's statistical efficiency can be much improved.
Both simulation studies and  real data analysis  demonstrate
the superior  advantage of our method.

\end{abstract}

\newpage

\csection{INTRODUCTION}

Real data analysis often runs into situations, where statistical inference is  too complicated to be analytically attractable. This could happen if the target statistics is a  complicated function of  sample moments (e.g., sample correlation coefficient). In this case, various bootstrap methods \citep{efron1990more,efron1994introduction} become practically appealing for their outstanding ability in automatic inference.
Automatic inference refers to the fact that important inference parameters (e.g., standard error) can be computed without knowing its analytical formula.
The resulting estimators  are  generally consistent \citep{van1996weak} and   could be  more accurate than those based
upon asymptotic approximation \citep{hall1994methodology}.

Traditional bootstrap (TB) methods are typically applied to datasets of  small sizes.
 In that case, computation is not an issue. This enables bootstrap methods to  be used  for automatic statistical inference. Unfortunately, such a standard paradigm becomes   problematic if  the datasets  are of  massive sizes.
In this case,  the computational  cost is no longer negligible. Instead,  it could be  rather challenging
even if one   single point estimate  needs to be  computed  based on the whole sample.
Accordingly a straightforward implementation of the traditional bootstrap methods becomes practically infeasible.
To mitigate this problem,  one possible solution  is to  utilize  the  parallel and distributed computing system.
However, this might lead to huge commuication   cost between different computer   nodes
\citep{kleiner2014scalable}.
Therefore, how to conduct statistically valid and computationally efficient  bootstrap  inference for massive datasets becomes a problem of great importance.

  Motivated by the need for an automatic
 statistical inference with  massive data, \cite{kleiner2014scalable} introduced  a new method called  as  Bag of Little Bootstraps
(BLB). For each bootstrap iteration,   a traditional bootstrap method should draw a  sample  of the same size as the original data. In case of massive data, this leads to
huge computational cost.
BLB replaces this  procedure  by two novel steps. In the first step, BLB obtain a first bootstrap sample with   size much smaller  than that of the  whole data.  In the second step, BLB obtains a second bootstrap sample by simple random sampling with replacement from the first  bootstrap sample. The size of the  second   bootstrap  sample  needs to be the same as that of the whole dataset.

 At the first glance, it seems that the  BLB  method should  lead  to  no less computational cost than  the traditional bootstrap  because the sample size involved in the second stage is as large as the whole dataset. However, the merit of BLB is that the second   bootstrap sample  is obtained from the first   bootstrap  sample. Since   the first bootstrap  sample size  is much smaller than that of the whole  dataset,   the second  bootstrap sample could be done by directly generating
 the frequency number for each data point in the first   bootstrap sample.
 The frequency number here refers to the number of times for a data point  in the first bootstrap sample  to be selected by the second step bootstrap sample.
  Accordingly, the target parameter can be computed in a weighted manner.
 The computational  cost of a BLB method becomes  the same order as that of the  first  bootstrap sample, which is much smaller than that of the TB methods.
On the other hand,  in theory,  the second   bootstrap sample is of the same size as the whole data.  Consequently, no analytical re-scalling is needed  \citep{kleiner2014scalable}.
 That makes BLB a fully automatic  method for statistical inference.

However,  the outstanding  performance  of BLB  relies  on three important    hyperparameters, which need to be  selected.
Specifically,  the  three hyperparameters   are,  respectively, (1)  the size of the first stage  bootstrap sample  $n$; (2) for a given first step bootstrap sample, the total number of bootstrap datasets $B$ needs to be generated in the second step; (3) the total  number of overall  Monte Carlo iterations $R$.
With unlimited computational resources, those  hyperparameters should be  as large as possible, because the larger they are, the  more accurate the resulting statistical inference is.
With limited computational budget, these three  hyperparameters need to  be carefully selected.  \cite{sengupta2016subsampled} have pointed out that under limited computational cost,  it remains unclear how to choose the optimal hyperparameters in BLB
that balances statistical accuracy and running time. In the meanwhile,  understanding such kind of  relationships  are  important as they  can have an empirical guidance for  hyperparamer selection.

In this work,  we  are seeking to find  the optimal balance between the statistical efficiency and computational cost.  To solve this problem,
we start with  the simplest statistic sample mean and its standard error (SE).  We next move on to more sophisticated statistics.
We start with sample mean mainly  for its simplicity. It leads to fruitful  and insightful theoretical  findings. These findings can be easily extended to more sophisticated statistics without much difficulty.
 We study SE because it  is an important parameter that  needs to be estimated for many important statistical inference. These important statistical inferences include, but are not limited to  confidence interval, hypotheses testing, and others.
 To estimate SE, one might  rely on asymptotic approximation. This is typically done by sophisticated Talylor's series approximation. Nevertheless, this method becomes challenging if the target statistic is too complicated to have an analytically  tractable  formula. In this case, automatic inference method (such as bootstrap) becomes practically appealing. This  amounts to use a method like BLB  to estimate the SE. The resulting estimation accuracy (about SE) should be closely related to  various  hyperparameters. Its relationship should be theoretically investigated.
 For theoretical  completeness, we have studied  the   estimation accuracy  of  four  popularly used  bootstrap  methods. They are, respectively, the traditional bootstrap  (TB, Efron, 1990), the bag of little bootstraps method (BLB, Kleiner et al., 2014), the $m$-out of-$n$ bootstrap  (SB, Bickel et al., 1997), and the subsampled double bootstrap (SDB, Sengupta et al., 2016).

 By a careful theoretical analysis, we find that the asymptotic efficiency of the subsampling estimate is closely related to its hyperparameters. The relationship is analytically simple and elegant.
  For illustration purpose, we consider the BLB method, as  it involves a total of three  hyperparameters (i.e., $n$, $R$, $B$).  For a given computational  platform, the time cost for BLB
 can be approximate by  $\beta_1 (nBR)+\beta_2 (nR)$ for some positive coefficients  $\beta_1$ and  $\beta_2$. Both the coefficients are computational platform specific and can be consistently estimated.  Our simulation experience suggests that this leads to fairly accurate
approximation; see subsection 4.3 for the details. Moreover,  the  details  of the hyperparameter selection approach are to be  introduced in subsection 3.2.
Once $\beta_1$ and $\beta_2$ are given,
we then  minimize $\mbox{MSE}(\hat{\mbox{SE}}^2_{C}) $
under the constraint $\beta_1 nBR+ \beta_2 nR  \leq C_{\max}$, where  $C_{\max}$ is the maximum  time cost we can bear.
 This leads to an optimal choice of the hyperparameters. More specifically for an arbitrarily specified hyperparameter set, we can improve it to be a new set of hyperparameters. The consequence is that no extra CPU time is needed but the resulting estimator's statistical efficiency can be much improved. Extensive numerical studies are conducted to demonstrate its performance.

  The rest of this article is organized as follows. We first introduce different bootstrap methods and then study their associated theoretical properties in subsection 3.1. The hyperparameter selection approach is presented in  subsection 3.2. Next, more general parameters and statistics are to be introduced in subsection 3.3.
   Extensive numerical studies are conducted in Section 4. Lastly, the article is concluded with a short discussion in Section 5.

\csection{BOOTSTRAP METHODS}

\csubsection{ Traditional Bootstraps}

We start with  the traditional bootstrap (TB) method.  Let  $X_1, X_2,\cdots,$ $X_N \in \mR^1$  be independent  and identically  distributed random variables with mean  $\mu$ and variance  $\sigma^2$.
For simplicity purpose, we start with the simple parameter $\mu$, which can be estimated by $\overline X =N^{-1}\sum_{i=1}^{N}X_i$. More complicated parameters and statistics are to be studied subsequently.
The estimation accuracy of $\overline X$  can be reflected by its standard deviation, which is also referred to as a Standard Error (SE).
Specifically,  the SE of  $\overline X$ is $\sigma/\sqrt{N}$ and it is  analytically simple. Consequently,  it can be estimated by
$\hat{\mbox{SE}}_{\tiny\mbox{A}} = \hat\sigma/\sqrt{N}$, where $\hat \sigma^2= \sum_{i=1}^{N} (X_i-\overline X)^2/N$.
This is an estimator obtained by analytical formula. We thus refer  to it as an AF estimator.
In the meanwhile, it can be estimated by a standard bootstrap  method as follows. For simplicity, we assume $X_i\in \mR^1$ is a scalar. However, the theory to be presented hereafter can be readily applied to multivariate  $X_i$s   without any difficulty.

Let   $\mS=\{X_1, \cdots,  X_N\}$ be the whole sample dataset and $B$  be the total number of bootstrap  datasets. Then, for any  $b=1,\cdots, B$,
 the $b$th bootstrap dataset is given by $\mB_{B}^{(b)}=\{X_{1,B}^{(b)}, X_{2,B}^{(b)}, \cdots, X_{N,B}^{(b)} \}$, where
 $X_{i,B}^{(b)}$s (for each $1\leq i\leq N$) are independently generated by the method of simple random sampling with replacement from the whole sample dataset $\mS$.
Based on $\mB_{B}^{(b)}$, the $b$th bootstrap sample mean can be calculated  as
$\overline X_B^{(b)}= N^{-1}\sum_{i=1}^{N} X_{i,B}^{(b)}$. Then, an estimate for SE$^2$ is given by
\beqr
\hat {\mbox{SE}}^2_{B}  = B^{-1} \sum_{b=1}^{B} \Big(\overline X_B^{(b)}- \overline X \Big)^2. \n
\eeqr
Conditional on  the whole sample dataset $\mS$ and assume appropriate regularity conditions, it can be    verified that   $\hat {\mbox{SE}}_{\tiny\mbox{B}}^2/ \hat {\mbox{SE}}_{\tiny\mbox{A}}^2 \rightarrow_p 1$,  as  $B\rightarrow \infty$; see Theorem 2 in Section 3 for the details.

 \csubsection{ Bag of Little Bootstraps}

 We next provide a brief review about the Bag of Little Bootstraps (BLB), which is a novel method  proposed by   \cite{kleiner2014scalable}.  The BLB method should be carried out by a number of  random replications.
  Here, we use $R$ to represent the total number of random replications. For each $1\leq r\leq R$, we should obtain a little bootstrap sample, which is denoted as  $\mB_{C}^{(r)}=\{X^{(r)}_{1,C},X^{(r)}_{2,C},\cdots, X^{(r)}_{n,C}\}$.
 where
 $X^{(r)}_{i,C}$s are independently
 generated by the method of simple random sampling with replacement from the whole sample  dataset $\mS$. It is remarkable that the size of $\mB_{C}^{(r)}$ is $n$, which is much smaller than the whole dataset $N$.

 In theory,  we should do a second stage bootstrap sampling of size $N$ from each $\mB_{C}^{(r)}$.  This   leads  to another  $B$  bootstrap samples given by
 $\mB_{C}^{(r,b)}= \{  X_{i,C}^{(r,b)}:   1\leq i \leq N\}$, where $ X_{i,C}^{(r,b)}$s are independently generated from $\mB_{C}^{(r)}$ by  simple random sampling with replacement.
Based on $\mB_{C}^{(r,b)}$, the target statistic $\overline X$ can be computed as
\beq
\overline X_C^{(r,b)}= N^{-1} \sum_{i=1}^N   X_{i,C}^{(r,b)} =  N^{-1}  \sum_{i=1}^n  X^{(r)}_{i,C} f_{i,C}^{(r,b)}. \label{star}
\eeq
 Here,  $ f_{i,C}^{(r,b)}$ is the sampling  frequency of $X_{i,C}^{(r)}\in \mB_{C}^{(r)}$.
 That is
 $ f_{i,C}^{(r,b)}=\sum_{j=1}^N  I(X_{i,C}^{(r,b)}= X_{j,C}^{(r)}$ and $I(\cdot)$ is an indicator function.
 Obviously, the random vector $ f_C^{(r,b)}=(f_{1,C}^{(r,b)},$ $ f_{2,C}^{(r,b)}, \cdots, f_{n,C}^{(r,b)})^\top\in \mR^n$ follows a multinomial distribution with parameter $N$ and $p$. Here,  $p=( 1/n,1/n,\cdots, 1/n)^\top\in \mR^n$.

 This leads to an interesting observation. In theory, a total of $N$  bootstrap samples
 in  $\mB_{C}^{(r,b)}$  needs to be generated. However, it is practically infeasible because $N$ is ultra large.
 In the meanwhile, by  \eqref{star}, we find that it is equivalent to  generate  the frequency vector $ f_C^{(r,b)}$ directly. The dimension of $ f_C^{(r,b)}$ is $n$, which is much smaller than $N$.  That makes the computational cost much cheaper.
Then,  with the help  of  $\overline X_C^{(r,b)}$, the target parameter SE$^2$ can be estimated by
 \beqr
\hat {\mbox{SE}}^2_{C}= \frac{1}{BR} \sum_{r=1}^R   \sum_{b=1}^{ B} \Big(\overline X_C^{(r,b)}- \overline X_C^{(r)}\Big)^2, \label{SE_c} \n
\eeqr
where $\overline X_C^{(r)} =n^{-1}\sum_{i=1}^{n}X^{(r)}_{i,C}$.
Conditional on the whole sample dataset $\mS$ and assume appropriate regularity conditions,   it can be verified  that  $\hat {\mbox{SE}}_{C}^2 / \hat {\mbox{SE}}_{A}^2\rightarrow_p 1$,  as $\min \{ B,R\}\rightarrow \infty$;
see Theorem 3 in Section  3 for the details.

\csubsection{ Subsampled   Bootstrap}

As one can see,  the BLB method  is closely related to another popularly used bootstrap method. That is  so-called ``$m$-out of-$N$'' bootstrap method \citep{bickel1997resampling}. For convenience,
we refer to it as a method of subsampled bootstrap (SB).

Let $B$ be the total number of bootstrap samples. For any $1\leq b\leq B$, we use $\mB_{D}^{(b)}=\{X_{i,D}^{(b)}: 1\leq i\leq n\}$ to denote the $b$th bootstrap sample with size $n$. Here,
$X_{i,D}^{(b)}$s are generated  independently  by  the simple random sampling with replacement from the whole sample dataset $\mS$.
 Based on $\mB_{D}^{(b)}$, the target statistic  $\overline X$
can be  computed as
$\overline X_D^{(b)}= n^{-1} \sum_{i=1}^n   X_{i,D}^{(b)}. $
Accordingly, the target inferential  parameter  SE$^2$  can be estimated by
\beqr
\hat {\mbox{SE}}^2_{D}= \Big(\frac{n}{N}\Big){ B}^{-1} \sum_{b=1}^{B} \Big(\overline X_D^{(b)}-\overline X\Big)^2. \label{SE_d} \n
\eeqr
Conditional on the whole sample dataset $\mS$,  one can verify that  $\hat {\mbox{SE}}_{D}^2 /\hat {\mbox{SE}}_{A}^2\rightarrow_p  1$,  as  $B\rightarrow \infty$; see Theorem 4 in Section 3 for the details.
Comparing   the formula of   $\hat {\mbox{SE}}^2_{D}$ with (for example) that of  $\hat {\mbox{SE}}^2_{B}$, we can find that a re-scaling factor  ($n/N$) is  needed  for $\hat {\mbox{SE}}^2_{D}$.
This re-scaling factor requires the knowledge of the convergence rate of the target estimator.  This makes the SDB method
 less automatic \citep{kleiner2014scalable,sengupta2016subsampled}.

\csubsection{ Subsampled Double Bootstrap}

Note that the  BLB method is also closely related to another interesting  bootstrap method for massive data analysis.
That is  the subsampled double bootstrap (SDB, Sengupta et al., 2016).  The implementation of the  SDB method  is similar to that of the  BLB method.

In the first stage,  SDB randomly draw  $R$ small subsets of the data.  For each $1\leq r\leq R$, the associated  little sample bootstrap subset is denoted as  $\mB_{E}^{(r)}=\{X^{(r)}_{1,E},X^{(r)}_{2,E},\cdots, X^{(r)}_{n,E}\}$,
 where
 $X^{(r)}_{i,E}$s are independently
 generated by the method of simple random sampling with replacement from the whole sample  dataset $\mS$.
 In the second stage, we only generate one bootstrap sample  from each subset $\mB_{E}^{(r)}$, which is denoted as $\mB_{E}^{(r,1)}=\{ X^{(r,1)}_{1,E}, X^{(r,1)}_{2,E},\cdots, X^{(r,1)}_{N,E}\}$.
 Here,  $X^{(r,1)}_{i,E}$s are  independently  generated from  $\mB_{E}^{(r)}$ by simple  random sampling with replacement.
 Based on $\mB_{E}^{(r,1)}$, the target  statistic  $\overline X$ can be  computed as
 \beq
\overline X_E^{(r,1)}= \frac{1}{N} \sum_{i=1}^N   X_{i,E}^{(r,1)} =  \frac{1}{N} \sum_{i=1}^n  X^{(r)}_{i,E} f_{i,E}^{(r,1)}. \label{star2}
\eeq
Here,  $ f_{i,E}^{(r,1)}$ is the frequency of $X_{i,E}^{(r)}\in \mB_{E}^{(r)}$.
That is  $ f_{i,E}^{(r,1)}=\sum_{j=1}^N I(X_{i,E}^{(r,1)}=X_{j,E}^{(r)}).$
 The random vector $f_E^{(r,1)}=(f_{1,E}^{(r,1)}, f_{2,E}^{(r,1)} \cdots, f_{n,E}^{(r,1)})^\top \in \mR^n$ follows the multinomial distribution with parameter $N$ and $p$, where  $p=( 1/n,1/n,\cdots, 1/n)^\top\in \mR^n$.
  Then, with the help of
  $\overline X_E^{(r,1)}$, the target parameter SE$^2$ can be estimated by
 \beqr
\hat {\mbox{SE}}^2_{E}= R^{-1} \sum_{r=1}^R    \Big(\overline X_E^{(r,1)}-\overline X_E^{(r)}\Big)^2, \label{SE_e} \n
\eeqr
where $\overline X_E^{(r)} =n^{-1}\sum_{i=1}^{n}X^{(r)}_{i,E}$.
Conditional on the whole sample dataset $\mS$ and assume  appropriate regularity  conditions,  it can be verified  that  $\hat {\mbox{SE}}_{E}^2 /\hat {\mbox{SE}}_{A}^2\rightarrow_p  1$,  as $R\rightarrow \infty$;
see Theorem 5 in Section 3 for the details.

\csection{ THEORETICAL PROPERTIES}
\csubsection{Theoretical Properties }

We next  study the theoretical properties for   $\hat{\mbox{SE}}^2_{A}$ to  $\hat{\mbox{SE}}^2_{E}$. Recall that $\hat{\mbox{SE}}_{A} = \hat\sigma/\sqrt{N}$,
 where $\hat \sigma^2= \sum_{i=1}^{N} (X_i-\overline X)^2/N$.
Define  $E(X_i-\mu)^4= \sigma_4$. We then have the following theorem.

\begin{theorem}
		For the  AF estimator,
	 we  have
\beqr
E(\hat{\mbox{SE}}^2_{A})=\frac{\sigma^2}{N}\Big(1-\frac{1}{N}\Big)
~\mbox{and}~
\var(\hat{\mbox{SE}}^2_{A})= \frac{\sigma_4 - \sigma^4}{N^3}\Big \{1 +o(1) \Big\}. \n
\eeqr
\end{theorem}
\noindent
By the above theorem result,  we can immediately obtain  the MSE for $\hat{\mbox{SE}}^2_{A}$, which can be expressed as
\beqr
\mbox{MSE}(\hat{\mbox{SE}}^2_{A}) =\frac{\sigma_4-\sigma^4}{N^3}\Big \{1 +o(1)\Big\}.
\label{MSE_a}
\eeqr

We next consider  the theoretical properties for   $\hat{\mbox{SE}}^2_{B}$. It is remarkable that conditional on $\mS$,
 $X^{(b)}_{i,B}$s  are independent and identically distributed. Specifically, we should have
 $P(X_{i,B}^{(b)}=X_j)=1/N$ for any $1 \leq j\leq N$.
Accordingly, we have $E(\overline X_B^{(b)}|\mS)= E(X_{i,B}^{(b)}|\mS)=\overline X$
and $\var(\overline X_B^{(b)}|\mS)= N^{-1}\var(X_{i,B}^{(b)}|\mS)=N^{-1} \hat\sigma^2.$
Recall that
$\hat {\mbox{SE}}^2_{B}  = B^{-1} \sum_{b=1}^{B} (\overline X_B^{(b)}- \overline X)^2. $ To study the theoretical properties
of  $\hat {\mbox{SE}}^2_{B}$, for  each $b=1\cdots, B$, we
 define $Y_B^{(b)}=(\overline X_B^{(b)} - \overline X)^2$ for convenience. Conditional on $\mS$,
$Y_B^{(b)}$s  are independent and identically distributed with $E(Y_B^{(b)}|\mS)=\var(\overline X_B^{(b)}|\mS)=\hat\sigma^2/N$.
We then obtain the following theorem.
\begin{theorem}
		For the TB estimator,
	 we have
\beqr
E(\hat{\mbox{SE}}^2_{B}) =\frac{\sigma^2}{N }\Big(1-\frac{1}{N}\Big) ~~\mbox{and}~~
\var(\hat{\mbox{SE}}^2_{B}) =\var (\hat{\mbox{SE}}^2_{A})\Big(\frac{2\sigma^4}{\sigma_4- \sigma^4}\cdot \frac{N}{B} +1 \Big)\Big \{1+o(1) \Big\}.\n
\eeqr
\end{theorem}
\noindent
From the above theorem, we can immediately obtain  the MSE for $\hat{\mbox{SE}}^2_{B}$, which can be expressed as
 \beqr
 \mbox{MSE}(\hat{\mbox{SE}}^2_{B}) &=& \var (\hat{\mbox{SE}}^2_{A})\Big(\frac{2\sigma^4}{\sigma_4- \sigma^4}\cdot \frac{N}{B} +1 \Big)\Big \{1+o(1) \Big\} + \frac{\sigma^4}{N^4}\n\\
 &=&(\sigma_4-\sigma^4) \Big(  \frac{2\sigma^4}{\sigma_4- \sigma^4}\cdot \frac{1 }{N^2 B}  + \frac{1 }{N^3} \Big) \Big\{1 + o(1)\Big\}.
 \label{MSE_b}
 \eeqr
This suggests  that
the bootstrap sampling size $B$ needs to be  the same/ larger order of $N$, if
$\mbox{MSE}(\hat{\mbox{SE}}^2_{B})$ wants to achieve the same convergence  rate with
$\mbox{MSE}(\hat{\mbox{SE}}^2_{A})$.

Recall that  $\overline X_C^{(r,b)}= N^{-1} \sum_{i=1}^N   X_{i,C}^{(r,b)} $, $\overline X_C^{(r)}= n^{-1} \sum_{i=1}^n   X^{(r)}_{i,C}, $
and  $\mB_{C}^{(r)}=\{   X^{(r)}_{i,C}:    1\leq  i \leq n\}$. Here, $\mB_{C}^{(r)}$ represent the sampling set in the $r$th replication.
It is remarkable that conditional on $\mS$ and $\mB_{C}^{(r)}$,
 $\overline X_C^{(r,b)}$s  are independent and identically distributed for $ 1\leq b\leq B$.
Accordingly, we have $E(\overline X_C^{(r,b)}|\mS,\mB_{C}^{(r)})=  \overline X_C^{(r)}$
and $\var(\overline X_C^{(r,b)}|\mS,\mB_{C}^{(r)})=  N^{-1} \hat\sigma_{r,C}^2,$
 where $\hat \sigma_{r,C}^2= n^{-1}\sum_{i=1}^{n} ( X^{(r)}_{i,C}-\overline X_C^{(r)})^2$.
  Since
$\hat {\mbox{SE}}^2_{C}= (BR)^{-1}\sum_{r=1}^R   \sum_{b=1}^{ B} (\overline X_C^{(r,b)}-\overline X_C^{(r)})^2,$
we next define ${Y_C^{(r,b)}}=(\overline X_C^{(r,b)}-\overline X_C^{(r)})^2$ for $b=1\cdots, B$. Then, conditional on $\mS$ and $\mB_{C}^{(r)}$,
${Y_C^{(r,b)}}$s  are independent and identically distributed with
\beqr
E\Big({Y_C^{(r,b)}}\Big|\mS,\mB_{C}^{(r)}\Big)=\var\Big(\overline X_C^{(r,b)}\Big|\mB_{C}^{(r)}\Big)=N^{-1}\hat\sigma_{r,C}^2.
\label{EY1}
\eeqr
  Consequently, we have the following theorem.
  \begin{theorem}
		For the BLB estimator,
	 we  have
\beqr
   E(\hat {\mbox{SE}}^2_{C})
   &=&  \frac{ \sigma^2}{N} (1-\frac{1}{n}) \Big\{1+ o(1)\Big\},
   \n\\
\var (\hat{\mbox{SE}}^2_{C})  &=&  \var (\hat{\mbox{SE}}^2_{A})\Big(  \frac{2\sigma^4}{\sigma_4- \sigma^4}\cdot\frac{ N} {RB}+\frac{N}{nR}+1\Big)\Big\{1+ o(1)\Big\}.\n
   \eeqr
\end{theorem}
\noindent
From the above theorem, we can immediately obtain  the MSE for $\hat{\mbox{SE}}^2_{C}$, which can be expressed as
\[ \mbox{MSE}(\hat{\mbox{SE}}^2_{C}) = \var (\hat{\mbox{SE}}^2_{A})\Big(  \frac{2\sigma^4}{\sigma_4- \sigma^4} \cdot \frac{ N} {RB}+\frac{N}{nR}+1\Big)\Big\{1+ o(1)\Big\} +\frac{\sigma^4}{N^2n^2}\]
\[=\frac{\sigma_4-\sigma^4}{N^2} \Big(  \frac{2\sigma^4}{\sigma_4- \sigma^4} \cdot  \frac{ 1} {RB}+\frac{1}{nR}  + \frac{\sigma^4}{\sigma_4- \sigma^4} \cdot \frac{1}{n^2} \Big)\Big\{1+ o(1)\Big\} +\frac{\sigma_4-\sigma^4}{N^3}\Big\{1+ o(1)\Big\}.
\]

 We further  study  the theoretical properties for $\hat{\mbox{SE}}^2_{D}$. Recall  that
 conditional on $\mS$,
 $X_{i,D}^{(b)}$s  are independent and identically distributed with
 $P(X_{i,D}^{(b)}=X_j)=1/N$ for any $1 \leq j\leq N$.
Accordingly, we have $E(\overline X_D^{(b)}|\mS)= E(X_{i,D}^{(b)}|\mS)=\overline X$
and $\var(\overline X_D^{(b)}|\mS)= n^{-1}\var(X_{i,D}^{(b)}|\mS)=n^{-1} \hat\sigma^2.$
Since  $\hat {\mbox{SE}}^2_{D}= n{ (NB)}^{-1} \sum_{b=1}^{B} (\overline X_D^{(b)}-\overline X)^2, $
we next define ${Y_D^{(b)}}=(\overline X_D^{(b)}- \overline X)^2$ for $b=1\cdots, B$. Conditional on $\mS$,
${Y_D^{(b)}}$s  are independent and identically distributed with $E({Y_D^{(b)}}|\mS)=\var(\overline X_D^{(b)}|\mS)=\hat\sigma^2/n$.
We further have the following theorem.
  \begin{theorem}
		For the SB estimator,
	 We  have  $E(\hat {\mbox{SE}}^2_{D})
   =N^{-1} (1-n^{-1} )\sigma^2$ and
\beqr
\var (\hat{\mbox{SE}}^2_{D}) =\var (\hat{\mbox{SE}}^2_{A})\Big(\frac{2\sigma^4}{\sigma_4-\sigma^4}\cdot\frac{N}{B}+1\Big) \Big\{1+o(1)\Big\}. \n
   \eeqr
\end{theorem}
Similarly, we can  obtain  the MSE for $\hat{\mbox{SE}}^2_{D}$, which can be expressed as
 \beqr
 \mbox{MSE}(\hat{\mbox{SE}}^2_{D}) &=& \var (\hat{\mbox{SE}}^2_{A})\Big(\frac{2\sigma^4}{\sigma_4-\sigma^4}\cdot\frac{N}{B}+1\Big)\Big \{1+o(1)\Big\}  +
 \frac{ \sigma^4}{N^2n^2} \n\\
 &=& \frac{\sigma^4}{N^2} \Big(\frac{2}{B}+  \frac{1}{n^2} +
 \frac{ \sigma_4-\sigma^4}{\sigma^4} \cdot \frac{1}{N}\Big)\Big \{1+o(1)\Big\}.
  \label{MSE_d}
 \eeqr

Recall that  $\overline X_E^{(r,1)}= N^{-1} \sum_{i=1}^N   X_{i,E}^{(r,1)} $.
It is remarkable that conditional on $\mS$,
 $\overline X_E^{(r,1)}$s  are independent and identically distributed for $ 1\leq r\leq R$.
Accordingly, we have $E(\overline X_E^{(r,1)}|\mS,\mB_{E}^{(r)})=  \overline X_E^{(r)}$
and $\var(\overline X_E^{(r,1)}|\mS,\mB_{E}^{(r)})=  N^{-1}\hat\sigma_{r,E}^2.$
  Since
$\hat {\mbox{SE}}^2_{E}= R^{-1}\sum_{r=1}^R    (\overline X_E^{(r,1)}-\overline X_E^{(r)})^2,$
we next define ${Y_E^{(r,1)}}=(\overline X_E^{(r,1)}-\overline X_E^{(r)})^2$ for $r=1\cdots, R$. Conditional on $\mS$ and $\mB_{E}^{(r)}$ ,
${Y_E^{(r,1)}}$s  are independent and identically distributed with
\beqr
E\Big({Y_E^{(r,1)}}\Big|\mS, \mB_{E}^{(r)}\Big)=\var\Big(\overline X_E^{(r,1)}\Big|\mS, \mB_{E}^{(r)}\Big)= \hat\sigma_{r,E}^2/N.
\label{YE1}
\eeqr
  Consequently, we have the following theorem.
  \begin{theorem}
		For the SDB estimator,
	 we  have
\beqr
   E(\hat {\mbox{SE}}^2_{E})
   &=&  \Big(\frac{ \sigma^2}{N}- \frac{ \sigma^2}{Nn}\Big) \Big\{1+ o(1)\Big\},
   \n\\
\var (\hat{\mbox{SE}}^2_{E}) &=& \var (\hat{\mbox{SE}}^2_{A})\Big(  \frac{2\sigma^4}{\sigma_4-\sigma^4}\cdot\frac{ N} {R}+1\Big)\Big\{1+ o(1)\Big\}.\n
   \eeqr
\end{theorem}
\noindent
From the above theorem, we can immediately obtain  the MSE for $\hat{\mbox{SE}}^2_{E}$, which can be expressed as
\beqr
 \mbox{MSE}(\hat{\mbox{SE}}^2_{E}) =\var (\hat{\mbox{SE}}^2_{A})\Big(  \frac{2\sigma^4}{\sigma_4-\sigma^4}\cdot\frac{ N} {R}+1\Big)\Big\{1+ o(1)\Big\}+
 \frac{ \sigma^4}{N^2n^2}\Big\{1+o(1)\Big\}\n\\
= N^{-2}\sigma^4 \Big( \frac{2}{\sigma_4-\sigma^4}\cdot \frac{ 1} {R} +\frac{1}{n^2}\Big)\Big\{1+ o(1)\Big\} +\frac{\sigma_4-\sigma^4}{N^3}\Big\{1+ o(1)\Big\}.
\label{MSE_e} \eeqr

\csubsection{Hyperparameter Selection for the BLB Method}

The fruitful theoretical results obtained in the previous subsections can be used to guide us to  search for the optimal hyperparameter specification. The objective here is to find  the   optimal hyperparameter specification  so that the resulting statistical efficiency (in terms of MSE) is minimal. For illustration purpose, we consider the BLB method only, because it represents the most complicated case here, as  it involves a total of three  hyperparameters (i.e., $n$, $R$, $B$).

The BLB method is mostly useful if the whole dataset is too large to be read into  a computer memory as a whole. In that case, the whole dataset has to be placed on a hard drive, but the computation  can only happen in  the memory. Accordingly, BLB needs  to repeatedly  sample $n$ data points  from the disk to the memory. Consequently, the time cost  for BLB mainly involves two parts. The first part is the sampling cost, which occurs while sample
 data from the disk to the memory.
There are a total of $R$ iterations and for each iteration there are $n$ data points need to be sampled. Thus, the associated time  cost should be the order of $O(nR)$.
This part is referred to as  the sampling cost.
For a given sampling iteration, once the $n$ data points have been   read into the memory,    one need to compute   $\overline X_C^{(r,b)}$ and  $\overline X_C^{(r)}$ first,
 which need $O(n)$ flopping operations. Furthermore, one needs to compute $\overline X_C^{(r,b)}$ for a total of $BR$ times. This leads to  $O(nBR)$ flopping operations.
Moreover, one needs to compute$(\overline X_C^{(r,b)}-\overline X_C^{(r)})^2$ for $BR$ times. This cost amounts to  $O(BR)$ flopping operations.  The total computational cost is given by $O(nBR)$ flopping operations.
This leads to the second part time cost, which is referred to as the  computational cost.
Combing the  sampling and computational  cost together, this leads to the total time cost as $O(nBR)+O(nR)$.
For a given computational  platform (e.g., a work station), we use the CPU time as a measure for the time cost.  It can be approximate by  $\beta_1 (nBR)+\beta_2 (nR)$ for some positive coefficients  $\beta_1$ and  $\beta_2$. Both the coefficients are computational platform specific.
Our simulation experience  suggests that  this leads to fairly accurate approximation. The simulation details are to be given in subsection 4.3.

Once $\beta_1$ and $\beta_2$ are given,
we then  minimize $\mbox{MSE}(\hat{\mbox{SE}}^2_{C}) $
under the constraint $\beta_1 nBR+ \beta_2 nR  \leq C_{\max}$, where  $C_{\max}$ is the maximum  time cost we can bear. Write  $c=\sigma^4(\sigma_4-\sigma^4)^{-1}$.
 This   amounts to find  the minimum value of
 $ 2c(RB)^{-1}+ (nR)^{-1}+c(n)^{-2} $ under the constraint  $\beta_1 n BR+ \beta_2 nR=C_{\max}$.
 To this end, let $(n,B,R)$ be an arbitrary specification such that  $\beta_1 nBR+ \beta_2 nR  = C_{\max}$. Then, due to
  Cauchy's inequality,
\beqr
&& \Big( 2c/(RB)+1/(nR)+ c /n^2 \Big) \Big(\beta_1  n BR+ \beta_2 nR\Big) \n\\
 = &&\Big(\frac{ 2c} {RB}+\frac{1}{nR} \Big) \Big(\beta_1  n BR+ \beta_2 nR \Big)+\frac{ c} {n^2} \Big(\beta_1  n BR+ \beta_2 nR \Big)
 \n \\
 \geq  && \Big(\sqrt {2 c \beta_1 }  \sqrt {n}  + \sqrt{\beta_2} \Big)^2 +  C_{\max} \frac{ c } {n^2}
= f(n),
 \label{c1}
\eeqr
where the function $f(n)$   is defined by the last equation in \eqref{c1}.
The  first  inequality  in \eqref{c1} becomes equality  if
 $ 2c/(\beta_1 nR^2 B^2)= 1/(\beta_2 n^2 R^2)$. When $n$ is fixed,
this  leads to
$B^*=\lfloor  (2c \beta_2/\beta_1) n )^{1/2} \rfloor, R^*= \lfloor C_{\max}/(\beta_1 n B^* +\beta_2 n)\rfloor,$
where  $ \lfloor s\rfloor$  stands for the largest integer  that no larger than $s$.
This suggests that the CPU time needed by  $(n, B^*, R^*) $ is
no more than that of  the  $(n, B, R) $.  However, the statistical efficiency (in terms of MSE) is likely to be better. Thus, $( B^*, R^*) $ can be viewed as the optimal specification when $n$ is fixed.

\csubsection{General Parameters and Statistics }

In this subsection, we study  parameters (statistics) more general than mean (sample mean).
 Without loss of generality, we still assume  $X_i\in \mR^1$ is a scalar with mean
 $E(X_i)= \mu\in \mR^1$. Next, we consider a more general parameter $\theta=g(\mu)$,
 where $g(\cdot)$ is a known, possibly complicated, but  sufficiently smooth function.
 We   can then estimate  $\theta$ by $\hat \theta =g(\overline X)$. Asymptotically,
 we have $\hat \theta-\theta= \dot g(\mu)(\overline X-\mu)\{1+o_p(1)\}$,
 where  $\dot g(\cdot)$ stands for the first order derivative of
 $g(\cdot)$.
  We then have $\sqrt N(\hat \theta-\theta)\rightarrow_d N(0, \dot g^2(\mu)\sigma^2)$.
  This means that the asymptotic SE$^2$  of  $\hat \theta$ is given by
 $ \mbox{SE}^{*2}=\dot g^2(\mu) \sigma^2/N. $
 Accordingly, a natural estimator for $ \mbox{SE}^{*2}$  is given by
 $\wh {\mbox{SE}}_A^{*2}=\dot g^2(\overline X) \hat \sigma^2/N, $
 which can be easily  compute if $\dot g^2(\mu)$ is analytically simple.
 The computational cost needed is about $O(N)$.
 However, in many cases,   $ \dot g(\mu)$ could be  rather complicated.
In this case, various automatic inference methods could be practically appealing. To this end, we next extend various  resampling methods (as defined in Sections  2 and 3) to this case.

 We start with the TB method.
  Accordingly, a natural estimator for $ \mbox{SE}^{*2}$  by TB method is given by
 \beqr
 {\wh {\mbox{SE}}^{*2}}_B &=& B^{-1} \sum_{b=1}^{B}\Big\{g(\overline X_B^{(b)})- g(\overline X)\Big\}^2. \label{SE_B*}
 \eeqr
 It can be further expressed as
 \beqr
  &=&  B^{-1} \sum_{b=1}^{B} \dot g^2(\overline X)\Big(\overline X_B^{(b)}-\overline X\Big)^2+o_p(N^{-1}) \n \\
 &=& \Bigg\{ B^{-1} \sum_{b=1}^{B} \dot g^2(\mu)\Big(\overline X_B^{(b)}-\overline X\Big)^2 \Bigg\} \Big\{1+o_p(1)\Big\}+o_p(N^{-1}) \n\\
&=& \dot g^2(\mu) {\wh {\mbox{SE}}^{2}}_B  \Big\{1+o_p(1)\Big\}. \n
\eeqr
This suggests that the asymptotic behavior of ${\wh {\mbox{SE}}^{*2}}_B$ is  largely determined by that of ${\wh {\mbox{SE}}^{2}}_B$. Under the same computational constraint
$C_{\max}N$, the best convergence rate can be achieved by  ${\wh {\mbox{SE}}^{*2}}_B$ should be the same as that of  ${\wh {\mbox{SE}}^{2}}_B$.

Similar analysis can be  conducted for BLB, SB and SDB  methods. The associated estimator  for     $ \mbox{SE}^{*2}$   are given by
 \beqr
{\wh {\mbox{SE}}^{*2}}_C&=&  (BR)^{-1} \sum_{r=1}^R   \sum_{b=1}^{ B} \Big\{g(\overline X_C^{(r,b)})- g(\overline X_C^{(r)})\Big\}^2= \dot g^2(\mu)\wh {\mbox{SE}}_C^2 \Big\{1+o_p(1)\Big\},\label{SE_C*}
\\
 {\wh {\mbox{SE}}^{*2}}_D&=& \Big(\frac{n}{N}\Big){ B}^{-1} \sum_{b=1}^{B} \Big\{g(\overline X_D^{(b)})-g(\overline X)\Big\}^2
=  \dot g^2(\mu)\wh {\mbox{SE}}_D^2 \Big\{1+o_p(1)\Big\}, \label{SE_D*}
\\
{\wh {\mbox{SE}}^{*2}}_E&=& R^{-1} \sum_{r=1}^R    \Big\{ g(\overline X_E^{(r,1)})-g(\overline X_E^{(r)})\Big\}^2
=  \dot g^2(\mu)\wh {\mbox{SE}}_E^2 \Big\{1+o_p(1)\Big\},  \label{SE_E*}
\eeqr
respectively.
We find  that the asymptotic behavior of ${\wh {\mbox{SE}}^{*2}}_C$, ${\wh {\mbox{SE}}^{*2}}_D$, and ${\wh {\mbox{SE}}^{*2}}_E$  are   largely determined by those  of ${\wh {\mbox{SE}}^{2}}_C$, ${\wh {\mbox{SE}}^{2}}_D$, and ${\wh {\mbox{SE}}^{2}}_E$. Under the same computational constraint,
the best convergence rate can be achieved by  ${\wh {\mbox{SE}}^{*2}}_C$, ${\wh {\mbox{SE}}^{*2}}_D$ and ${\wh {\mbox{SE}}^{*2}}_E$  should be the same as those  of   ${\wh {\mbox{SE}}^{2}}_C$, ${\wh {\mbox{SE}}^{2}}_D$ and ${\wh {\mbox{SE}}^{2}}_E$.

\csection{ NUMERICAL STUDY}

\csubsection{The consistency  for SE$^2$}

The objective of this subsection is to numerically confirm whether the various subsampling methods (e.g., BLB) studied in this work  can indeed  estimate the true  SE$^2$ consistently.
As we discussed before, even though our primary  theory is developed for sample mean,    it can be readily applied  for more complicated statistics.
 For illustration purpose, we can consider here a slightly  more complicated statistic, that is  sample correlation coefficient \citep{ross2017introductory}. Specifically, we  have observations $(X_i, Y_i)s$  identically and  independently generated from  a bivariate normal distribution with mean 0, covariance  $(1,\rho;\rho,1)\in\mR^{2\times2}$ and $\rho=0.5$, where $i=1,\cdots, N$.
Then, the  sample correlation coefficient is given by
\beqr
\hat \rho= \frac{\sum_{i=1}^{N}(X_i-\bar X)(Y_i-\bar Y)}{\sqrt{\sum_{i=1}^{N}(X_i-\bar X)^2}\sqrt{ \sum_{i=1}^{N}(Y_i-\bar Y)^2}}, \label{4.1}
\eeqr
where  $\bar X=N^{-1}\sum_{i=1}^n X_i$ and  $\bar Y=N^{-1}\sum_{i=1}^{N} Y_i$.
Let $M$ be the total number of simulation replications.
  Write $\hat \rho^{(m)}$
 be the estimator for $\rho$  obtained in the $m$th simulation replication for $1 \leq m\leq M$.
  Then, its true squared standard error (SE$^2$) can be consistently estimated by
 $ \mbox{SE}^{*2}= M^{-1} \sum_{m=1}^{M}(\hat \rho^{(m)}- \rho)^2. $

Let $\hat{\mbox{SE}}^{2(m)}$ be one particular SE$^2$ estimate obtained in the $m$ th simulation replication. For example, it could be  the TB estimator.
Next, define $\gamma^{(m)}= \hat{\mbox{SE}}^{2(m)}/{\mbox{SE}}^{*2}$. If $\hat{\mbox{SE}}^{2(m)}$ is a consistent estimator for SE$^2=\var(\hat \rho)$, we should expect $\gamma^{(m)}$ to be very close to 1.
Accordingly. the sample mean (MEAN) and sample standard deviation (SD) are computed and reported in Table 1.
Various parameter settings are considered. Specifically,
we fix  $N =10^5$,      $n\in \{\lfloor N^{0.5}\rfloor,  \lfloor N^{0.6}\rfloor, \lfloor N^{0.7}\rfloor\}$.
Set  $B\times R \in \delta$, where $\delta =\{25,50,75,100,125,150,175,200,225,250,275,300\}$.
Since for TB and SB methods, they  do not involve the hyperparameter $R$, we then set $B \in \delta$ for these two methods. Similarly, for SDR method, we set  $R \in \delta$.
 For each  fixed parameter setup, a total of $M=1,000$ random replications have been conducted.
The detailed simulation results are summarized in Table 1.

As we can see from this table, the sample mean of $\gamma^{(m)}$ for $1 \leq m\leq M$ is generally all across to 1  for different bootstrap  methods, which suggests the correctness of the  SE$^{*2}$  estimators  defined in \eqref{SE_B*} to  \eqref{SE_E*}. Moreover, when  $n$ and $\delta$ are fixed,  the associated  SD values    are   more or less the same for different methods. Besides, the SD values are consistently decreasing  as   either $\delta$  or $n$ increases.
Take TB estimator  for illustration  purpose, the  associated SD values for $\gamma$ decrease from 0.291 to 0.085, when $(\delta,n)$ increases from  $(25,\lfloor N^{0.5}\rfloor)$ to  $(300, \lfloor N^{0.7}\rfloor)$.

\csubsection{The  Mean Squared Error}

The objective of this subsection is to numerically confirm whether the analytical formula for  MSE($\hat{\mbox{SE}}^2$) defined in subsection 3.1 are indeed correct. Here,
$\hat{\mbox{SE}}^2$ stands for one particular estimator for  $\mbox{SE}^2$. For example, it could be  the SB or SDB estimator. Accordingly, the analytical formula to be evaluated is given by \eqref{MSE_d} and  \eqref{MSE_e}. Since the analytical formula are available only for sample mean.  Thus, the statistic studied  in this subsection is sample mean.  Here, two different distributions are considered for $X_i$s ($1\leq i\leq N$). They are, respectively,     standard normal distribution (Normal)  or   centralized standard exponential distribution (Exp).
 Let $ \hat {\mbox{SE}}^{2(m)}$ be one particular  SE$^2$ estimate (e.g., the  BLB estimator) obtained in the $m$ th  ($1 \leq m\leq M$) simulation replication. Its true mean squared error (MSE)  can  then be estimated as
 \beqr
\hat{\mbox{MSE}}= M^{-1}\sum_{m=1}^M \Big(\hat {\mbox{SE}}^{2(m)}-{\mbox{SE}}^{*2}\Big)^2
\label{MSE_cp}
 \eeqr
 where $\mbox{SE}^{*2} = M^{-1}\sum_{m=1}^M(\overline X^{(m)}-\overline X)^2$ and
 $\overline X^{(m)}$ is the $\overline X$ estimator  obtained in the $m$ th ($1\leq m\leq M$) simulation replication.

 Let $\mbox{MSE}^{*}$ be the oriented  MSE value computed according to our theory.
 For example, by \eqref{MSE_a} we should be able to compute the theoretical MSE  for
  $\hat {\mbox{SE}}_A^2$. We then evaluate   the difference between $\hat {\mbox{MSE}}$  and $\mbox{MSE}^{*}$ by a ratio  $\kappa= \mbox{MSE}^{*}/\hat{\mbox{MSE}}$. It should be close to 1 if our theory is correct.
The detailed simulation results are
reported in Table \ref{tab:t2}.  In this table, we fix $N=10^5$ and  various parameter settings are considered.
Specifically, we consider  $n\in \{\lfloor N^{0.4}\rfloor,  \lfloor N^{0.5}\rfloor, \lfloor N^{0.6}\rfloor\}$, $B\in \{25,50\}$, and $R\in \{25,50\}$.
For each  fixed parameter setup, a total of $M=1,000$ random replications have been conducted.  The detailed simulation results are summarized in Table 2.
Note that,  the ratio  $\kappa$  for AF estimator  does not contain any hyperparameter, thus the value for it would not change. Similarly, the TB estimator only involves the hrperparameter $B$, thus the  values of  $\kappa$ for TB method  would  not change with $R$ and $n$.
Overall, we find that
  the  values of  $\kappa$   are   very close to  one  across different simulation scenarios.
This result   suggests that  correctness of the  theoretical  formula for  MSE$ ( \hat {\mbox{SE}}^2_{A})$ to MSE$( \hat {\mbox{SE}}^2_{E})$, which have been given in  subsection 3.1.

\csubsection{Hyperparameter Selection}

The objective of this subsection is to evaluate  how the selection   of the hyperparameters (e.g., $B,R$) would   affect the performance of BLB. The performance of other subsampling methods are similar and less complicated. We then omitted for short.
In that subsection, the BLB method is  implemented with
repeatedly reading dataset from disk, instead of reading from memory.
Specifically,
consider  the sample correlation coefficient defined in \eqref{4.1},
where  $(X_i,Y_i)$s  are generated  from a  bivariate  normal distribution with mean 0 and covariate $(1,0.5;0.5,1)$.
 We fix $N=5\times 10^5$, $n=5000$, and
consider  $B\times R= \delta \times 10^3$,  where $\delta\in \{1,2,3,4\}$.
Based on the value of   $\delta$, we generate 16 different $(B,R)$   combinations to verify the  performance of our hyperparameter selection method while using BLB to  estimate SE$^2$.

 We first consider how to estimate the coefficient $\beta_0=(\beta_1,\beta_2)$, which is used to approximate the actual CPU time and find the optimal specification of   $( B^*, R^*) $.  Here,  the CPU time represents  the time to conduct  BLB  with fixed $(B,R)$.  The  16  current  $(B,R)$   combinations are used  to  estimate $\beta_0$. To enlarge the variation of $B$ and $R$, we further    generate another 10 combinations of  $( B, R)$   from  $\lfloor   10\times U(2,100)\rfloor$ and  $\lfloor 10\times U(1,20)\rfloor$, respectively.
For each  $(B,R)$ combination,  a total of  $20$  random replications have been conducted  and   we recorded the median CPU time. This leads to a total of 26 time records and they are treated as responses.
 We then use the corresponding $n\times B\times R$ and $n\times R$ as covariates, which are  called as $nBR$ and $nR$ hereafter for short.
This  leads to  the   coefficient estimator as $\hat \beta_1=2.342 \times 10^{-7}$ and  $\hat \beta_2=1.076\times 10^{-4}$.  The resulting R.Squared is as high as 98\%, which  suggests that the approximation is quite accurate. The demanded CPU time can then be estimated by  $C_{\max}= \hat \beta_1  (nBR)+\hat \beta_2  (nR)$. Based on $C_{\max}$, we further obtain the  optimal specification as
 $ B^*=\lfloor  (2c \hat \beta_2/\hat\beta_1) n )^{1/2} \rfloor$ and  $R^*= \lfloor C_{\max}/(\hat \beta_1 n B^* +\hat\beta_2 n)\rfloor.$

Next, both $(n,B,R)$ and $(n, B^*,R^*)$  specifications are used to estimate  SE$^2$.
For each fixed parameter setup, a total of $M=200$ random replications have been conducted.
The resulting estimators' statistical efficiency are then evaluated in a similar manner as in Section 4.1. The resulting    mean squared errors (reported by the median value)  are denoted by MSE$_a$ and MSE$_b$, respectively.
Their  log-transformed form are denoted by $\log(\mbox{MSE}_a)$ and $\log(\mbox{MSE}_b)$ accordingly.
 The associated CPU times are denoted by  Time$_a$ and Time$_b$, which are reported in seconds. For comparison purpose, two ratios  $\mbox{MSE}_b/\mbox{MSE}_a$ and  $\mbox{Time}_b/ \mbox{Time}_a$ are also  computed.

The detailed simulation results are summarized in Table 3.   We find that  the ratio  $\mbox{Time}_b/ \mbox{Time}_a$  are generally close or smaller  than 1.  This suggests  that the CPU time consumed by $(n,B,R)$ and $(n, B^*,  R^*)$ are comparable. On the other hand,
we find that   $\mbox{MSE}_b/\mbox{MSE}_a$ are all well below 100\%. It could be as small as 10\%, e.g., when $( B,R)=(10, 100)$, the associated $\mbox{MSE}_b/ \mbox{MSE}_a=9.7\%$. In most cases,  it stays well below 80\%. This implies that  the estimators provided by $(n, B^*,R^*)$
 are statistically more efficient that  those provided by $(n,B,R)$. The  averaged improving  margin across all simulation cases  is as large as 38.5\%.

\csubsection{ Real  Data Analysis}

In this subsection, we consider a  U.S. Airline Dataset  for illustration purpose.
This is a dataset about detailed flight information. We take the data in the year of 2008 for illutration purpose, which contains $N=1.01\times 10^6$  records. It is publicly  available at http://stat-computing.org/dataexpo/2009.
The objective of the study  is to evaluate how the selection of the hyperparameters (e.g., $B,R$) would affect the  SE performance.
  Our focus is to  estimate   the  SE$^2$  for sample correlation between distance and airline arrival time.
Note that the original   sample
correlation  between distance and airline arrival time equals to $0.975$.  To enhance the variability between the two variables, we further add a standard error term  $N(0,1)$ to  the airline arrival time, which leads to  the sample
correlation
  as   $0.574$.
 Because we have no knowledge of  the true  correlation $\rho$  between distance and airline arrival time.  We then  report the
 $\wh {\mbox{SE}}^{2}$   instead  of MSE  for the sample correlation coefficient $\hat\rho$.

   Since BLB is the most complicated subsampling method, in this subsection, we still use BLB for illustration purpose.
  Specifically, we  generated 16 different $(B,R)$ combinations based on $B\times R= \delta \times 10^3$ with $\delta\in \{1,2,3,4\}$.
   Using the same coefficient estimation method introduced   in subsection 4.3. We further obtained  the specified optimal $(B^*,R^*)$ accordingly.
For each hyperparameter
setup, the experiment is randomly replicated for $M=200$ times.
  In each  replication, the  associated  $\wh {\mbox{SE}}^{2}$ for $\hat \rho$    can be calculated via BLB accordingly.  Specifically, in the $m$ th random replication, it is denoted as   $\wh {\mbox{SE}}_C^{2(m)}$
  for each $1 \leq m\leq M$.
   The resulting   median value of  $\wh {\mbox{SE}}^{2}$   for
   $(n,B,R)$ and $(n,B^*,R^*)$ are denoted by MedSE$_a$ and  MedSE$_b$, respectively.
   To measure the stability of   $\wh {\mbox{SE}}^{2}$, we further  calculate  the standard deviation of  it under the M random replications.
 Let SD$_a$ and SD$_b$ be the standard deviation of  $\wh {\mbox{SE}}^{2}$ for
   $(n,B,R)$ and $(n,B^*,R^*)$, respectively.
Similar with subsection 4.3, the
CPU times are recorded to  control the time cost budget. Specifically, let  Time$_a$ and Time$_b$ be the associated
CPU time  for    $(n,B,R)$ and $(n,B^*,R^*)$, respectively.
 For comparison purpose,  we then report  the ratios   $\mbox{MedSE}_b/\mbox{MedSE}_a$,
  SD$_b$/SD$_a$,   and  Time$_b$/Time$_a$ in Table 4.

The detailed  results are summarized in Table 4. First, we find that  the ratio  $\mbox{Time}_b/ \mbox{Time}_a$  are generally close or smaller  than 1.  This suggests  that the CPU time consumed by $(n,B,R)$ and $(n, B^*,  R^*)$ are comparable.  Second,  we find that   $\mbox{MedSE}_b/\mbox{MedSE}_a$ are all very close to 1 for all cases. This implies that both $\mbox{MedSE}_a$ and $\mbox{MedSE}_b$ are consistent with each other.
On the other hand,
we find that   $\mbox{SD}_b/\mbox{SD}_a$ are all well below 100\%. It could be as small as 20\%, e.g., when $( B,R)=(10, 100)$, the associated $\mbox{SD}_b/\mbox{SD}_a=19.9\%$. In all  cases,  it stays well below 80\%. This implies that  the estimators provided by $(n, B^*,R^*)$
 are more stable  that  those provided by $(n,B,R)$.

\csection{ CONCLUDING REMARKS}

In this article we propose a hyperparameter selection approach which can be applied   for subsampling methods. These sampling methods are first applied to  estimate the standard error (SE) for sample mean.
We study SE because it is an important parameter that needs to be estimated for many
important statistical inference.
 It then  move  on to more sophisticated statistics in subsection 3.3.
For theoretical completeness, we have studied the associated  theoretical properties of the  SE estimators for  AF, TB, BLB, SB, and SDB, respectively.
 Given these  theoretical findings,  we then proposed a methodology  to do hyperparameter selection.
 Since BLB is the most complicated subsampling method, we have used it for illustration purpose.
Similar approaches  can be readily  applied to  other subsampling methods, e.g., SB and SDB.

\csection{ Acknowledgments}

Yingying Ma's research is partially supported by National Natural Science Foundation of China (No.11801022).
Hansheng Wang's research is partially supported by National Natural Science Foundation of China (No.11831008, 11525101, 71532001).
It is also supported in part by China's National Key Research Special Program (No. 2016YFC0207704).

\begin{landscape}

\begin{table}[t]
\bc\emph{}
\caption{\label{tab:t1} Detailed simulation results  are reported  for $\gamma$ in  Section 4.1 with $N=10^5$. The  SD for  $\gamma$ are also  recorded in parentheses.}
{\small
\begin{tabular}{c|lc|lc|lc|lc}
\hline
\hline
 \multicolumn{1}{c|}{ } & \multicolumn{2}{|c}{TB}  & \multicolumn{2}{|c}{BLB} & \multicolumn{2}{|c}{SB}& \multicolumn{2}{|c}{SDB}\\
   Parameter & Parameter&MEAN( SD)  &  Parameter&MEAN( SD)  &Parameter&MEAN( SD)  & Parameter&MEAN( SD)   \\
\hline
   $n=\lfloor N^{0.5} \rfloor$   &    B=25  &      1.038 (0.291) &     B=5,  R=5  &      1.005 (0.290)  &       B=25 &     1.042 (0.308)  &   R=25  &     1.013 (0.287)   \\
                               &    B=50  &       1.023 (0.207) &    B=5, R=10 &      1.026  (0.206) &       B=50 &     1.048  (0.206) &    R=50 &     1.023 (0.218)  \\
                               &     B=75 &       1.029 (0.170) &    B=15, R=5 &      1.016  (0.184) &       B=75 &     1.035 (0.166)  &     R= 75 &     1.019 (0.167)  \\
                               &    B=100 &     1.027 (0.140) &      B=10, R=10 &      1.021  (0.155) &       B=100 &    1.034  (0.146)  &   R=100 &     1.022 (0.147)  \\
 &       &       & &&&&\\
 $n=\lfloor N^{0.6} \rfloor$     &    B=125  &     1.022 (0.134) &    B=5,  R=25  &     1.032 (0.130) &    B=125 &     1.028 (0.134) &   R=125 &     1.024 (0.131)  \\
                               &    B=150 &     1.021 (0.118) &     B=15,  R=10 &     1.031 (0.128) &    B=150 &     1.022 (0.112) &   R=150 &     1.027 (0.115)  \\
                               &    B=175 &     1.024 (0.111) &     B=25,  R=7  &     1.025 (0.118) &    B=175 &     1.031 (0.111) &   R=175 &     1.029 (0.105) \\
                               &    B=200 &     1.028 (0.105) &     B=10,  R=20 &     1.020 (0.102) &    B=200 &     1.030 (0.107) &   R=200 &     1.026 (0.107)  \\
 &       &       & &&&&\\
$n=\lfloor N^{0.7} \rfloor$       &B=225 &     1.022  (0.098) &       B=15,  R=15 &     1.032 (0.099) &    B=225 &     1.030 (0.098) &     R=225 &     1.024 (0.096)   \\
                                &B=250 &     1.026  (0.090) &       B=10,  R=25 &     1.029 (0.098) &     B=250 &     1.032 (0.093) &     R=250 &     1.027 (0.092)  \\
                                &B=275 &     1.028  (0.087) &       B=25,  R=11 &     1.026 (0.093) &     B=275 &     1.024 (0.088) &     R=275 &     1.026 (0.092)   \\
                                &B=300 &     1.029  (0.085) &       B=10,  R=30 &     1.026 (0.086) &     B=300 &     1.032 (0.083) &     R=300 &     1.027 (0.086) \\
   \hline
   \end{tabular}}
\ec
\end{table}

\begin{table}[t]
\bc\emph{}
\caption{\label{tab:t2} Detailed simulation results  are reported  for $\kappa$ in  Section 4.2 with $N=10^5$. }
{\small
\begin{tabular}{ccc|ccccc|ccccc}
\hline
\hline
 \multicolumn{3}{c|}{Parameter} & \multicolumn{5}{|c}{Normal}  & \multicolumn{5}{|c}{Exp}\\
   $n$ & $B$&   $R$&AF   &  TB&BLB   &   SB&SDB&AF     &  TB&BLB   &   SB&SDB\\
  \hline
   $ \lfloor N^{0.4} \rfloor$&    25 &    25 &     1.007   &     0.982 &     0.979 &     1.004 &     1.002 &     1.055 &     1.003 &     0.969 &     0.944 &     1.116  \\
                            &    25 &    50 &        1.007 &     0.982 &     0.914 &     1.004 &     1.010 &     1.055 &     1.003 &     1.001 &     0.944 &     1.061  \\
                             &   50  &    25 &   1.007 &     0.927 &     0.975 &     1.020 &     1.002 &     1.055 &     1.006 &     0.929 &     0.976 &     1.116   \\
                             &    50 &    50 &     1.007   &     0.927   &     0.987   &      1.020  &       1.010 &       1.055 &       1.006 &       1.023 &       0.976 &     1.061   \\

    &       &       & &&&&\\
     $ \lfloor N^{0.5} \rfloor$ &   25 &   25 &       1.007 &     0.982 &     1.042 &     1.075 &     1.000 &     1.055 &     1.003 &     0.991 &     0.964 &     1.069  \\
                                  &    25 &    50 &     1.007 &     0.982 &     0.982 &     1.075 &     1.014 &     1.055 &     1.003 &     1.028 &     0.964 &     1.032   \\
                                  &    50 &    25 &     1.007 &     0.927 &     0.937 &     1.014 &     1.000 &     1.055 &     1.006 &     0.962 &     1.013 &     1.069   \\
                                   &    50 &    50 &     1.007 &     0.927 &     0.950 &     1.014 &     1.014 &     1.055 &     1.006 &     0.946 &     1.013 &     1.032   \\

  &       &       & &&&&\\
  $ \lfloor N^{0.6} \rfloor$ &    25 &    25 &     1.007 &     0.982 &     0.946 &     0.877 &     0.932 &     1.055 &     1.003 &     1.044 &     0.967 &     0.951  \\
                               &    25 &    50 &     1.007 &     0.982 &     1.003 &     0.877 &     1.030 &     1.055 &     1.003 &     0.975 &     0.967 &     0.977   \\
                                  &    50 &    25 &     1.007 &     0.927 &     1.059 &     0.991 &     0.932 &     1.055 &     1.006 &     0.965 &     1.121 &     0.951   \\
                                   &    50 &    50 &     1.007 &     0.927 &     1.082 &     0.991 &     1.030 &     1.055 &     1.006 &     1.049 &     1.121 &     0.977   \\
                         \hline
   \end{tabular}}
\ec
\end{table}

\begin{table}[!h]
\bc\emph{}
\caption{\label{tab:t3} Detailed  simulation results  are reported  for Section 4.3 with $N=5\times 10^5$ and $n=5000$.}
\vspace{0.15cm}
{\small
\begin{tabular}{ccc|cc|ccc|ccc}
\hline
\hline
 \multicolumn{3}{c|}{Setting I} &  \multicolumn{2}{c|}{Setting II} & \multicolumn{3}{|c}{MSE Performance}  &  \multicolumn{3}{|c}{CPU Time}\\
 $BR(\times 10^3)$& $B$& $R$&    $B^*$& $R^*$&  $\log(\mbox{MSE}_a)$& $\log(\mbox{MSE}_b)$& MSE$_b$/MSE$_a$ & Time$_a$&Time$_b$ & Time$_b$/Time$_a$ \\
\hline
1&  100 &    10 &  1515 &     2 &   -33.484 &   -33.685 &     0.818 &     8.362 &     4.786 &     0.572  \\
 &  50 &    20 &  1515 &     5 &   -33.659 &   -34.651 &     0.371 &    12.317 &    10.981 &     0.891   \\
 &  20 &    50 &  1515 &    12 &   -33.592 &   -35.576 &     0.138 &    26.709 &    27.013 &     1.011   \\
  & 10 &   100 &  1515 &    23 &   -33.496 &   -35.829 &     0.097 &    49.304 &    51.323 &     1.041   \\
&   &&&&&&&&&\\
2  &100 &    20 &  1515 &     5 &   -34.332 &   -34.732 &     0.670 &    17.112 &    11.409 &     0.667   \\
  & 50 &    40 &  1515&    10 &   -34.158 &   -35.354 &     0.302 &    26.255 &    22.125 &     0.843   \\
  & 25 &    80 &  1515 &    19 &   -34.105 &   -35.845 &     0.175 &    44.727 &    41.897 &     0.937   \\
  & 10 &   200 & 1515 &    47 &   -34.158 &   -36.166 &     0.134 &    99.914 &   102.899 &     1.030   \\
&   &&&&&&&&&\\
 3 &100 &    30 &  1515 &     8 &   -34.505 &   -35.184 &     0.507 &    26.483 &    18.336 &     0.692   \\
  & 75 &    40 &  1515 &    10 &   -34.692 &   -35.441 &     0.473 &    31.459 &    22.653 &     0.720   \\
  & 50 &    60 &  1515 &    15 &   -34.671 &   -35.592 &     0.398 &    40.848 &    33.448 &     0.819   \\
  & 25 &   120 & 1515 &    29&   -34.568 &   -35.828 &     0.284 &    68.062 &    64.730 &     0.951   \\
&   &&&&&&&&&\\
4  &100 &    40 &  1515 &    11 &   -34.864 &   -35.271 &     0.666 &    35.481 &    25.258 &     0.712   \\
  & 50 &    80 &  1515 &    20 &   -34.971 &   -35.862 &     0.410 &    54.585 &    44.874 &     0.822   \\
  & 40 &   100&  1515 &    25&   -35.126 &   -35.847 &     0.487 &    63.572 &    55.229 &     0.869   \\
  & 20 &   200 &  1515 &    48 &   -34.854 &   -36.236 &     0.251 &   108.775 &   105.261 &     0.968 \\
\hline
  \end{tabular}}
\ec
\end{table}

\begin{table}[!h]
\bc\emph{}
\caption{\label{tab:t4} Detailed  results  are reported  for the  Airline Data with
 $N=1.01\times 10^6$  and $n=5000$.}
\vspace{0.15cm}
{\small
\begin{tabular}{ccc|cc|cc|ccc}
\hline
\hline
 \multicolumn{3}{c|}{Setting I} &  \multicolumn{2}{c|}{Setting II} & \multicolumn{2}{|c}{MedSE}  &  \multicolumn{3}{|c}{CPU Time}\\
 $BR(\times 10^3)$& $B$& $R$&    $B^*$& $R^*$&MedSE$_b$/MedSE$_a$& SD$_b$/SD$_a$&Time$_a$&Time$_b$ & Time$_b$/Time$_a$ \\
\hline
 1&100 &    10 &  1895 &     3 &     0.999 &     0.717 &    11.680 &     8.936 &     0.765   \\
 & 50 &    20 &  1895 &     5 &     1.007 &     0.507 &    19.740 &    14.533 &     0.736   \\
 & 20 &    50 &  1895 &    14 &     0.995 &     0.361 &    42.512 &    39.622 &     0.932   \\
 & 10 &   100 &  1895 &    27 &     1.003 &     0.199 &    80.443 &    75.841 &     0.943   \\
 &   &&&&&&&&\\
2 &100 &    20 &  1895 &     6 &     0.996 &     0.658 &    23.897 &    17.300 &     0.724   \\
 & 50 &    40 &  1895 &    11 &     1.005 &     0.459 &    40.162 &    31.282 &     0.779   \\
 & 25 &    80&  1895 &    22 &     1.001 &     0.322 &    70.489 &    61.872 &     0.878 \\
&  10 &   200 &  1895 &    55 &     1.004 &     0.233 &   160.895 &   153.872 &     0.956   \\
 &   &&&&&&&&\\
3 &  100 &    30&  1895 &     9 &     1.001 &     0.697 &    36.061 &    25.634 &     0.711   \\
  &  75&    40 &  1895 &    12 &     0.999 &     0.591 &    45.321 &    34.053 &     0.751   \\
 &   50 &    60 &  1895 &    17 &     0.998 &     0.567 &    60.520 &    47.971 &     0.793   \\
 &   25 &   120 &  1895 &    34 &     1.001 &     0.354 &   106.135 &    95.358 &     0.898   \\
 &   &&&&&&&&\\
4 &  100 &    40 &  1895 &    12 &     1.001 &     0.736 &    48.309 &    33.988 &     0.704   \\
 &   50&    80 &  1895 &    23 &     1.004 &     0.505 &    80.832 &    64.703 &     0.800   \\
 &   40 &   100 &  1895 &    29 &     0.999 &     0.455 &    95.977 &    81.399 &     0.848   \\
 &   20 &   200 &  1895 &    56 &     0.999 &     0.331 &   172.232 &   156.615 &     0.909 \\
\hline
  \end{tabular}}
\ec
\end{table}

\end{landscape}

\end{document}